\documentclass[conference]{IEEEtran}

\usepackage[letterpaper, left=0.680in, right=0.680in, bottom=1.0in, top=0.75in]{geometry}

\usepackage{ifpdf}
\ifCLASSINFOpdf
\usepackage[pdftex]{graphicx}

\DeclareGraphicsExtensions{.pdf,.jpeg,.png}
\else
\usepackage[dvips]{graphicx}
\DeclareGraphicsExtensions{.pdf}
\usepackage[caption=false, font=footnotesize]{subfig}
\fi

\usepackage{xcolor,soul,framed}
\usepackage[noadjust]{cite}

\usepackage{xcolor,soul,framed} 
\usepackage[noadjust]{cite}

\colorlet{shadecolor}{yellow}

\usepackage[pdftex]{graphicx}
\graphicspath{{../pdf/}{../jpeg/}}
\DeclareGraphicsExtensions{.pdf,.jpeg,.png}

\usepackage{amssymb,mathtools}
\usepackage{array}
\usepackage{mdwmath}
\usepackage{mdwtab}
\usepackage{eqparbox}
\usepackage{url}

\usepackage[ruled,vlined,linesnumbered]{algorithm2e}
\usepackage{enumitem} 
\usepackage[mathcal]{euscript} 
\usepackage{tabularx} 
\usepackage{multirow}
\usepackage{booktabs} 
\usepackage{makecell} 
\usepackage{aligned-overset} 
\usepackage{url}
\usepackage{multicol} 
\usepackage{balance}
\usepackage{bbm}
\usepackage{amsmath}
\usepackage{amsthm}

\usepackage{pifont}

\graphicspath{{./Figures/}}

\newtheorem{Lemma}{Lemma}

\newtheorem{Theorem}{Theorem}

\renewcommand{\vec}[1]{\boldsymbol{\mathrm{#1}}}

\newcommand{\s}{\mathrm{S}}

\newcommand{\tx}{\textnormal{S}}
\newcommand{\rx}{\textnormal{D}}
\newcommand{\IRS}{\textnormal{I}}
\newcommand{\ue}{\textnormal{D}}

\newcommand{\SINR}{\Gamma}

\newcommand{\gain}{\mathbb{G}}

\newcommand{\icd}{{(i)}}
\newcommand{\rfl}{{(s)}}

\newcommand{\pl}{\ell}

\newcommand{\tsqrt}[1]{{\textstyle{\sqrt{#1}}}}

\begin{document}
   \title{User Association Optimization for IRS-aided Terahertz Networks: A Matching Theory Approach}
  \author{\IEEEauthorblockN{
            Muddasir Rahim\IEEEauthorrefmark{1}, Thanh Luan Nguyen\IEEEauthorrefmark{1},
			Georges~Kaddoum\IEEEauthorrefmark{1}$^,$\IEEEauthorrefmark{3}, 
			and Tri~Nhu~Do\IEEEauthorrefmark{2},
			}
		\IEEEauthorblockA{
		\IEEEauthorrefmark{1}
		Department of Electrical Engineering, \'{E}cole de Technologie Sup\'{e}rieure (\'{E}TS), Universit\'{e} du Qu\'{e}bec, Montr\'{e}al, QC, Canada, \\Emails: muddasir.rahim.1@ens.etsmtl.ca,
            thanh-luan.nguyen.1@ens.etsmtl.ca, georges.kaddoum@etsmtl.ca}
   \IEEEauthorrefmark{3} Artificial Intelligence \& Cyber Systems Research Center, Lebanese American University \\
  \IEEEauthorrefmark{2} Department of Electrical Engineering, Polytechnique Montreal, QC, Canada, Email: tri-nhu.do@polymtl.ca

	}

\maketitle
\begin{abstract}
  Terahertz (THz) communication is a promising technology for future wireless communications, offering data rates of up to several terabits-per-second (Tbps). However, the range of THz band communications is often limited by high pathloss and molecular absorption. To overcome these challenges, this paper proposes intelligent reconfigurable surfaces (IRSs) to enhance THz communication systems. Specifically, we introduce an angle-based trigonometric channel model to evaluate the effectiveness of IRS-aided THz networks. Additionally, to maximize the sum rate, we formulate the source-IRS-destination matching problem, which is a mixed-integer nonlinear programming (MINLP) problem. To solve this non-deterministic polynomial-time hard (NP-hard) problem, the paper proposes a Gale-Shapley-based solution that obtains stable matches between sources and IRSs, as well as between destinations and IRSs in the first and second sub-problems, respectively.
\end{abstract}

\section{INTRODUCTION}\label{introduction}

The exponential growth in the number of connected devices and multimedia applications has dramatically increased the demand for large bandwidth (BW) and high data rate transmissions~\cite{rappaport2019wireless}. One of the vital service targets in sixth-generation (6G) is ultra-high-speed of up to 1 terabits-per-second (Tbps)~\cite{chowdhury20206g}. In this context, the terahertz (THz) band is considered a promising candidate for enabling such applications in 6G wireless networks~\cite{wan2021terahertz}. The THz band has many advantages, but establishing a reliable transmission link at THz frequencies is not easy. The reason is that there are strong atmospheric attenuations, extremely high free-space losses, and the line-of-sight (LOS) channel is highly sensitive to blockage effects. The consequences of these factors may include a reduction in service coverage and communication range for THz communication networks.

Recently, intelligent reconfigurable surfaces (IRSs) have become a promising technology to tackle the above problem of THz networks \cite{DoTCOMM2021}. IRSs are made of a massive number of small passive and metamaterial-based reconfigurable elements, and an IRS can manipulate the phase and/or the amplitude of incident signals to reflect the signals in the desired directions. 
Specifically, the use of IRSs is useful when the LOS channel is blocked or has weak received signal power since it is possible to provide additional transmission links by utilizing the reflecting elements of an IRS~\cite{di2020reconfigurable}. However, IRSs may have limited computing capability to support signal processing, especially when heterogeneous IRS deployment in a given network area is considered. In such a scenario, user pairing is essential for maximizing the IRS's efficiency. In addition, to maximize the sum rate through multiple IRS-aided networks, a fast and efficient IRS scheduling policy is vital.

Existing works on IRS-aided networks mostly focus on the phase shift (PS) matrix with passive beamforming at the IRS to maximize the signal-to-noise ratio (SNR), sum rate, secrecy rate, or energy efficiency, without considering IRS selection. There are few studies that addressed the problem of IRS-user scheduling in multiple-input-single-output (MISO) networks.
Y. Fang \textit{et al.} proposed IRS selection strategies for multi-IRS-aided wireless networks in which the number of elements of each IRS can be arbitrarily set~\cite{fang2020optimum}. The performance analysis was carried out assuming that the magnitudes of the channel coefficients associated with different IRSs are independent and identically distributed (i.i.d.) random variables. 
W. Mei \textit{et al.} proposed an IRS selection policy that maximizes the end-to-end (e2e) SNR for multi-IRS-aided networks~\cite{mei2020cooperative}. In this context, the authors only considered the impact of pathloss (PL) and ignored channel fading effects. 
S. Zhang \textit{et al.} demonstrated the effect of distributed and centralized IRS deployment schemes on the capacity region of multi-IRS-aided wireless networks~\cite{zhang2020intelligent}. According to the authors, in distributed IRS deployments, the channels associated with each distributed IRS are subject to i.i.d. Rayleigh fading.
I. Yildirim \textit{et al.} considered multi-IRS-aided wireless networks for both outdoor and indoor communications, where the direct channel between a source and a destination is blocked~\cite{yildirim2020modeling}. A low-complexity IRS selection scheme that uses the IRS with the highest SNR for communication is proposed in this study.
\subsubsection*{\textbf{Contributions}}\label{contributions}
In contrast to the aforementioned related works, this paper considers the source (S)-IRS-destination (D) scheduling problem in IRS-aided green wireless communication systems as a matching problem. The main contributions of this work are summarized as 1) In this paper, we derive the channel model for IRS-aided THz networks without direct channels. We also propose a new angle-based trigonometric PL model and compare it to existing models to validate its correctness. 2) To the best of our knowledge, this is the first paper to study the S-IRS-D association problem in IRS-aided THz networks for green communication systems. In order to maximize the sum rate, we formulate the association problem as a three-dimensional (3D) matching problem. Moreover, the 3D objective function is a mixed-integer nonlinear programming (MINLP) problem, which is challenging to solve. 3) It is proved that the proposed matching-based algorithm converges to a stable matching and terminates after a finite number of iterations. Numerical results demonstrate that the proposed matching-based association can significantly improve the sum rate compared to the greedy search (GS) and random assignment (RA). Additionally, the sum rate of the proposed method is compatible with the sum rate of the ES method.
\section{Description of the IRS-assisted THz System} \label{model}
\begin{figure}[t]
 \centering
 \includegraphics[width=0.75\linewidth]{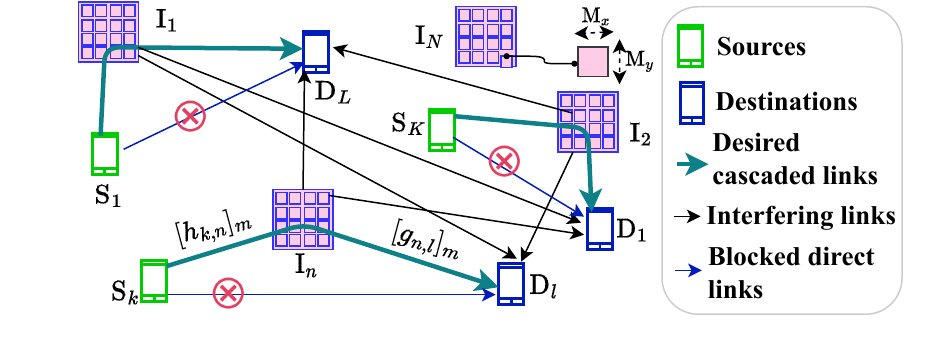}
 \caption{Illustration of the considered IRS-aided THz communication network.}
 \label{m1}
 \end{figure}

We consider IRS-assisted THz networks for green communication systems, where ${N}$ distributed IRSs are used to assist the communication between $K$ ($K \le N$) sources and $L$ destinations, with a single antenna each. In addition, each IRS is equipped with $M_n$ reflecting elements. The area of each IRS element is $A = M_x \times M_y$, where $M_x$ and $M_y$ denote the length of the horizontal and vertical sides, respectively. Henceforth, we use S, D, and I as acronyms for the source, destination, and IRS, respectively. In addition, we assume all direct links are blocked due to severe shadowing by obstacles or human bodies in the environment~\cite{wan2021terahertz}, and destinations are served through IRS cascaded links, as shown in Fig.~\ref{m1}.
\subsubsection{\textbf
{Channel Modeling}}
Assuming that $\tx_k$ transmits a signal $x_k$ with transmission power $P_{k}$, where $k \in {\cal K} = \{1,\dots, K\}$. We focus on the propagation model for cascaded links.
Let $\vec{h}_{k,n}{ \in \mathbb{R}}^{M \times 1}$ denotes the channel vector from $\tx_k$ to $\IRS_n$, where $n \in {\cal N} = \{1,\dots, N\}$. In this context, the individual channel from $\tx_k$ to $\IRS_{n,m}$ can be expressed as~\cite{DovelosICC2021} 
\begin{align}
    [\vec{h}_{k,n}]_m = h_{k,n,m} = \tsqrt{\pl_{\tx_k,\IRS_{n,m}}} e^{ - j \frac{2 \pi}{\lambda} d_{\tx_k,\IRS_{n,m}}},
\end{align}
where $\kappa \triangleq \frac{2 \pi}{\lambda}$ is the \textit{wavenumber}, $\lambda \triangleq \frac{c}{f_c}$ $[m]$ denotes the wavelength, $d_{\tx_k,\IRS_{n,m}}$ denotes the distance from $\tx_k$ to $\IRS_{n,m}$, and $\pl_{\tx_k,\IRS_{n,m}}$ represents the PL in $\tx_k$ to $\IRS_{n,m}$ link. Let $\vec{g}_{n,l} \in \mathbb{R}^{M \times 1}$ denotes the channel gain vector of the $\IRS_n$ to $\rx_l$ links and the individual channel from $\IRS_{n,m}$ to $\rx_l$ can be expressed as~\cite{DovelosICC2021}
\begin{align}
   [\vec{g}_{n,l}]_m = g_{n,m,l} = \tsqrt{\pl_{\IRS_{n,m},\rx_l}} e^{ - j \frac{2 \pi}{\lambda} d_{\IRS_{n,m},\rx_l}} ,
\end{align}
where $d_{\IRS_{n,m},\rx_l}$ denotes the distance from $\IRS_{n,m}$ to $\rx_l$ and $\pl_{\IRS_{n,m},\rx_l}$ represents the PL in the $\IRS_{n,m}$ to $\rx_l$ link. 
 Furthermore, based on the existing PL models for THz frequencies~\cite{DovelosICC2021, OzcanTVT2021}, we propose a new PL model that facilitates our network analysis.
\begin{Lemma} \label{lemma_gain}
Denoting ${ \gain_{\IRS_{n,m}}(\vec{r}^{\icd}, \vec{r}^{\rfl}) \triangleq  \gain_{\IRS_{n,m}}(-\vec{r}^{\icd}) \gain_{\IRS_{n,m}}(\vec{r}^{\rfl}) }$ as  cascaded gain of the $m^{th}$ element of $\IRS_n$, the angle-based trigonometric model of ${\gain_{\IRS_{n,m}}(\vec{r}^{\icd}, \vec{r}^{\rfl})}$ is formulated as 
\begin{align} \label{gain}
    \gain_{\IRS_{n,m}}(\vec{r}^{\icd}, \vec{r}^{\rfl}) 
&=  \bigg( 
        \frac{4 \pi A}{\lambda^2} 
        \eta_{\IRS_{n,m}}(-\vec{r}^{\icd}, \vec{r}^{\rfl})
    \bigg)^2, \\
\eta_{\IRS_{n,m}}(-\vec{r}^{\icd}, \vec{r}^{\rfl})
&\triangleq
    \sqrt{ \cos^2 \psi^{\icd} (\cos^2 \phi^{\rfl} \cos^2 \psi^{\rfl} +\sin^2 \phi^{\rfl})}. \nonumber
\end{align}
\end{Lemma}
\begin{IEEEproof}
Considering that the incident wave from $\tx_k$ is linearly polarized along the $x$-axis, 
its electric field (E-field) is presented as ${\vec{E}^{\icd}_{k} \!=\! E_{k}^\icd e^{-j \kappa (y\sin \theta^{\icd}_{k} - z\cos \theta^{\icd}_{k})} \vec{e}_{x}}$ \cite{DovelosICC2021}, where $E_{k}^\icd$ [V/m] denotes the amplitude of the incident E-field and $\theta^{\icd}_{k}$ [rad] denotes the incidence angle.
We rewrite the E-field as $\vec{E}^{\icd}_{k} = (E^{\icd}_{x, k}, E^{\icd}_{y, k}, E^{\icd}_{z, k}) $. Since ${\vec{e}_{x} = (1, 0,  0)}$, we obtain
\begin{align}
E^{\icd}_{x, k} = E_{k}^\icd
    e^{-j \kappa (y\sin \theta^{\icd}_{k} - z\cos \theta^{\icd}_{k})},
~E^{\icd}_{y, k} = E^{\icd}_{z, k} = 0.
\end{align}
Using Faraday equation, i.e., ${\vec{\nabla} \times \vec{E}^{\icd}_{k} = -j \omega \vec{B}^{\icd}_{k}}$ with ``$\times$'' being the cross product operator, we can determine the incident magnetic B-field ${\vec{B}^{\icd}_{k} \triangleq (B^{\icd}_{x, k}, B^{\icd}_{y, k}, B^{\icd}_{z, k})}$. 
    Here, the left-hand side of the Faraday equation is obtained as
\begin{align}
\vec{\nabla} \times \vec{E}^{\icd}_{k}
= \bigg(
    0, \frac{\partial E^{\icd}_{x, k}}{\partial z}, - \frac{\partial E^{\icd}_{x, k}}{\partial y}
\bigg),
\end{align}
where ${\vec{e}_{y} = (0, 1, 0)}$ and ${\vec{e}_{z} = (0, 0, 1)}$.
The partial derivative of $E^{\icd}_{x, k}$ with respect to $z$ and $y$ are obtained as ${j \kappa \cos\theta^{\icd}_k E^{\icd}_{x, k}}$ and ${-j \kappa \sin\theta^{\icd}_k E^{\icd}_{x, k}}$, respectively. 
    Hence, comparing the right-hand side of the Faraday equation, we obtain ${B^{\icd}_{x, k} = 0}$, ${B^{\icd}_{y, k} = - \frac{\kappa}{\omega} \cos \theta^{\icd}_k E^{\icd}_{x, k}}$, and ${B^{\icd}_{z, k} = -\frac{\kappa}{\omega} \sin \theta^{\icd}_k E^{\icd}_{x, k}}$. 
Hence, the magnetic B-field components of the incident wave is obtained~as
\begin{align}
\vec{B}^{\icd}_{k}
    = - \frac{E^{\icd}_k}{\eta} \Big[ \vec{e}_{y} \cos\theta^{\icd}_{k} + \vec{e}_{z} \sin \theta^{\icd}_{k} \Big] ^{-j \kappa (y\sin \theta^{\icd}_{k} - z\cos \theta^{\icd}_{k})},
\end{align}
where $\eta = {\omega} / {\kappa}$. Following the analysis in \cite[Section 11.3.2]{BalanisWiley2012} and \cite{DovelosICC2021}, the corresponding power density of the scattered E-field at $\rx_l$ is obtained as
\begin{align}
S^{\rfl}_{\tx_k, \IRS_n, \rx_l} &= 
    \frac{P_k \gain_{\tx_k}}{4 \pi}
    \bigg( \frac{\frac{A}{\lambda}}{d_{\tx_k, \IRS_{n,m}} d_{\IRS_{n,m}, \rx_l}} \bigg)^2
\nonumber\\
    &\qquad\qquad\quad\times
    F(\psi_{\tx_k, \IRS_{n,m}},\phi_{ \IRS_{n,m}, \rx_l },\psi_{ \IRS_{n,m}, \rx_l }).
\end{align}
Hence, considering the molecular absorption loss in THz communication and the receive aperture $\frac{\gain_{\rx_l} \lambda^2}{4\pi}$, the received signal at $\rx_l$ can be expressed as
\begin{align} \label{eq_rx_signal_v1}
y  = \sum_{m=1}^{M_n} 
    &\sqrt{ \frac{\gain_{\rx_l} \lambda^2}{4\pi} S^{\rfl}_{\tx_k, \IRS_n, \rx_l} } \kappa_{n, m} 
    \sqrt{e^{-\kappa_{\sf abs}(f_c)(d_{\tx_k,\IRS_{n,m}}+d_{\IRS_{n,m},\rx_l})}} 
    \nonumber \\
    & \quad \times 
    e^{ - j \frac{2 \pi}{\lambda} (d_{\tx_k,\IRS_{n,m}}+d_{\IRS_{n,m},\rx_l})} 
    e^{ j\theta_{n,m} } x_k + n_{\rx_l},
 \end{align}  
where 
    $F(x,y,z)= \cos^2 x (\cos^2 y \cos^2 z +\sin^2 y)$, $\kappa_{n,m} \in (0,1]$ and $\theta_{n,m} \in [0, 2\pi)$ represent the amplitude and the PS of $\IRS_{n,m}$, respectively. Moreover,
    $\kappa_{\sf abs}(f)$ $[\textnormal{m}^{-1}]$ is the molecular absorption coefficient at the frequency of interest $f$ [Hz].
For $f \in [275, 400]$ GHz, $\kappa_{\sf abs}(f)$ can be determined as \cite{TarboushTVT2021}
\begin{align}
\kappa_{\sf abs}(f)
    =   y_1(f, \mu_{\textnormal{H}_2 \textnormal{O}})
        +   y_2(f, \mu_{\textnormal{H}_2 \textnormal{O}})
        +   g(f),
\end{align}
where $y_1(f, \mu_{\textnormal{H}_2 \textnormal{O}})$, and $y_2(f, \mu_{\textnormal{H}_2 \textnormal{O}})$, $g(f)$ are given by \cite[Eq. (31), Eq. (32), Eq. (33)]{TarboushTVT2021}, respectively, $\mu_{\textnormal{H}_2 \textnormal{O}} = \frac{\phi}{100} \frac{p^\ast_w(T, p)}{p}$ is the volume mixing ratio of water vapor, $\phi \in [0, 100]$ is the relative humidity, and
    $\frac{\phi}{100} p^\ast_w(T, p)$ is the partial pressure of water vapor at temperature $T$ [K] and pressure $p$ [hPa]. 
Specifically, $p^\ast_w(T, p)$ is given by
\begin{align}
p^\ast_w(T, p) = 6.1121 \bigg(1.0007+ \frac{3.46}{10^6} p\bigg)
    e^{ \frac{17.502 (T-273.15)}{(T-32.18)} }.
\end{align}

Henceforth, we consider the standard atmospheric condition, where $T = 296$ K, $p = 1013.25$ hPa and $\phi = 50$, denoting $50\%$ humidity.
\color{black}
Furthermore, the received signal corresponding to~\cite{OzcanTVT2021}, can be expressed as
\begin{align} \label{eq_rx_signal_v2}
y  = \sum_{m=1}^{M_n} &\frac{\sqrt{P_{k}A^2 \gain_{\tx_k} \gain_{\IRS_{n,m}}(-\vec{r}_{\tx_k, \IRS{_n,m}})}}{4 \pi d_{\tx_k,\IRS_{n,m}} } \sqrt{e^{-\kappa_{\sf abs}(f_c)d_{\tx_k,\IRS_{n,m}}}}\nonumber\\ 
&\times 
\frac{\sqrt{A^2 \gain_{\rx_l} \gain_{\IRS_{n,m}}(\vec{r}_{\IRS_{n,m}, \rx_l})}}{4 \pi  d_{\IRS_{n,m},\rx_l}} \sqrt{e^{-\kappa_{\sf abs}(f_c)d_{\IRS_{n,m},\rx_l}}} 
\nonumber \\
& \quad \times 
e^{ - j \frac{2 \pi}{\lambda} (d_{\tx_k,\IRS_{n,m}}+d_{\IRS_{n,m},\rx_l})} 
e^{ j\theta_{n,m} } x_k + n_{\rx_l}.
\end{align}
Comparing~\eqref{eq_rx_signal_v1} and~\eqref{eq_rx_signal_v2}, we can formulate the gain of the $m^{th}$ element of $\IRS_n$ as \eqref{gain}.
This completes the proof of Lemma~\ref{lemma_gain}.
\end{IEEEproof}
 \begin{figure}[t]
 \centering
 \includegraphics[width=\linewidth]{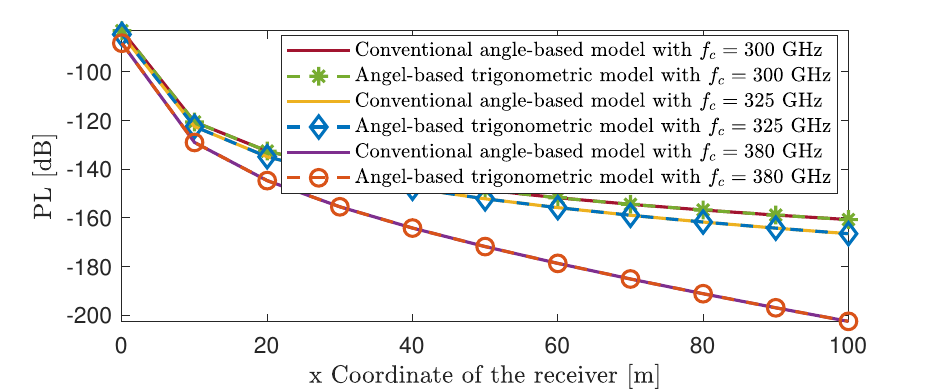}
 \caption{Comparison between the conventional angel-based and the proposed angel-based trigonometric PL models for different frequencies.}
 \label{pl_difmodel}
 \end{figure}
Invoking Lemma~\ref{lemma_gain}, the PL of the cascaded link is the combination of the PLs over the source to IRS and IRS to Rx links, which can be expressed as
\begin{align}
\pl_{\tx_k,\IRS_{n,m},\rx_l} 
    &= \frac{ \gain_{\tx_k} \gain_{\IRS_{n,m}}(-\vec{r}_{\tx_k, \IRS_{n,m}}) }{ (4\pi/\lambda)^2 }
    \frac{ e^{-\kappa_{\sf abs}(f_c) d_{\tx_k, \IRS_{n,m}}} }{ (d_{\tx_k, \IRS_{n,m}})^2 } \nonumber \\
    &\quad\times
    \frac{ \gain_{\IRS_{n,m}}(\vec{r}_{\IRS_{n,m}, \rx_l}) \gain_{\rx_l} }{ (4\pi/\lambda)^2 }
    \frac{ e^{-\kappa_{\sf abs}(f_c) d_{\IRS_{n,m},\rx_l}} }{ (d_{\IRS_{n,m},\rx_l})^2 },
\end{align}
\color{black}
where $\gain_{\IRS_{n,m}}(-\vec{r}_{\tx_k, \IRS_{n,m}})$ denotes the gains of $\IRS_{n,m}$ in the incident direction, and $\gain_{\IRS_{n,m}}(\vec{r}_{\IRS_{n,m}, \rx_l})$ are the gains of $\IRS_{n,m}$ in the direction $\vec{r}_{\IRS_{n,m}, \rx_l}$ \cite{DovelosICC2021}.
\begin{figure*} 
\normalsize
\begin{align} \label{eq_sinr_e2e1}
&\SINR_{\tx_k, \IRS_n, \rx_l} = 
    \frac{ 
    \left|
        \sum\limits_{m=1}^{M_n}
        \frac{ \sqrt{P_{k} 
        \gain_{\tx_k} 
        \gain_{\IRS_{n,m}}(-\vec{r}_{\tx_k, \IRS_{n,m}}) 
        \gain_{\IRS_{n,m}}(\vec{r}_{\IRS_{n,m}, \rx_l}) 
        \gain_{\rx_l} } }{ d_{\tx_k, \IRS_{n,m}} d_{\IRS_{n,m}, \rx_l} 
        e^{ \kappa_{\sf abs}(f_c) (d_{\tx_k, \IRS_{n,m}} + d_{\IRS_{n,m}, \rx_l})/2 } }e^{ - j \frac{2 \pi}{\lambda} (d_{\tx_k, \IRS_{n,m}}+d_{\IRS_{n,m}, \rx_l}){+j \theta_{n,m}} }
    \right|^2 }{ 
    \left|
        \sum\limits_{j\ne k}^K 
        \sum\limits_{i=1}^N 
        \sum\limits_{m=1}^{M_i}
        \frac{\sqrt{P_j 
        \gain_{\tx_j} 
        \gain_{\IRS_{i,m}}(\vec{r}_{\tx_j, \IRS_{i,m}}) 
        \gain_{\IRS_{i,m}}(\vec{r}_{\IRS_{i,m}, \rx_l}) 
        \gain_{\rx_l} 
        }}{ d_{\tx_j,\IRS_{i,m}} d_{\IRS_{i,m}, \rx_l} e^{ \kappa_{\sf abs}(f_c) (d_{\tx_j, \IRS_{i,m}} + d_{\IRS_{i,m}, \rx_l})/2 }  } e^{ -j \frac{2 \pi}{\lambda} (d_{\tx_j,\IRS_{i,m}} + d_{\IRS_{i,m}, \rx_l}){+j \theta_{i,m}}}
    \right|^2 +\sigma^2 },
\end{align}
\hrule
\end{figure*}
The PL assuming the conventional angel-based and the proposed angel-based trigonometric models with a single $\tx$/$\rx$ for different frequencies is demonstrated in Fig.~\ref{pl_difmodel}. The SINR at $\rx_l$, $\SINR_{\tx_k, \IRS_n, \rx_l}$, of the desired signal from $\tx_k$ and $\IRS_n$ is given in~\eqref{eq_sinr_e2e1},
where $\sigma^2$ is the AWGN noise power at each user.
In our system, we only take into account signals reflected by an IRS one time because the signals reflected by the IRS two times or more are weak and can be neglected~\cite{noh2022cell}.
\section{Problem Formulation}\label{pform}
As stated earlier, our primary goal is S-I-D scheduling to maximize the sum rate. For our system model, the achievable rate can be written as
\begin{align}\label{rateknl}
     R_{\tx_k,\IRS_n,\rx_l} = \log_2 (1+\SINR_{\tx_k,\IRS_n,\rx_l} ).
\end{align}
\subsubsection{\textbf{Problem Formulation}}
Considering the SINR as a function of $[\vec{\Omega}]_{\tx_k,\IRS_n,\rx_l}$,
where $\vec{\Omega}$ is the 3D association matrix.
Therefore, the objective is to find the optimal association matrix $[\vec{\Omega}]_{\tx_k,\IRS_n,\rx_l}$ that maximizes the sum-rate, expressed as
\begin{subequations}\label{eq_opt_prob}
\begin{flalign}
\vec{P} \text{ : } 
\underset{[\vec{\Omega}]_{\tx_k, \IRS_n, \rx_l}}{\textnormal{maximize}}
&\hspace{5pt}   \sum_{k=1}^K   R_{\tx_k, \IRS_n, \rx_l}, \label{eq_optprob} &\\
   \textnormal{subject~to} 
&\hspace{5pt}  \textstyle\sum_{k=1}^K [\vec{\Omega}]_{\tx_k, \IRS_n, \rx_l} = 1, 
    \forall{n} \in {\cal N}, \forall{l} \in {\cal L},\!\!\!\!
\label{eq_original_prob_constraint_1} &\\
{}&\hspace{5pt} \textstyle\sum_{n=1}^N [\vec{\Omega}]_{\tx_k,\IRS_n,\rx_l} = 1, 
    \forall{k} \in {\cal K}, \forall{l} \in {\cal L},\!\!\!\!
\label{eq_original_prob_constraint_2} &\\
{}&\hspace{5pt} \textstyle\sum_{l=1}^L [\vec{\Omega}]_{\tx_k, \IRS_n, \rx_l} = 1, 
    \forall{n} \in {\cal N}, \forall{l} \in {\cal L},\!\!\!
\label{eq_original_prob_constraint_3} &\\
{}&\hspace{5pt} [\vec{\Omega}]_{\tx_k, \IRS_n, \rx_l} \in \{0, 1\}, 
    \forall k \in {\cal K}, \forall n \in {\cal N}, \forall l \in {\cal L}, &
\label{eq_original_prob_constraint_4}
\end{flalign}
\end{subequations}
where the objective in~\eqref{eq_optprob} is the sum rate of the network given the pairing $[\vec{\Omega}]_{\tx_k, \IRS_n, \rx_l}$; constraint~\eqref{eq_original_prob_constraint_1} ensures that each S will be paired to one I and D only; constraint~\eqref{eq_original_prob_constraint_2} guarantees that each I is allocated to one S and D; constraint~\eqref{eq_original_prob_constraint_3} make sure that each D is paired with one I and S. Thus, constraints~\eqref{eq_original_prob_constraint_1}, \eqref{eq_original_prob_constraint_2}, and \eqref{eq_original_prob_constraint_3} ensure a one-to-one S-I-D matching. Additionally,~\eqref{eq_original_prob_constraint_4} enforces that $[\vec{\Omega}]_{\tx_i,\IRS_n,\rx_j}$ is either 1, i.e., the IRS is allocated, or 0, i.e., the IRS is free.
The formulated problem is an MINLP problem because it includes multiple integer variables, e.g., $[\vec{\Omega}]_{\tx_k,\IRS_n,\rx_l} \in \{0,1\}$, and the objective function is nonlinear. Additionally, it can be observed that the optimization problem in~\eqref{eq_opt_prob} includes the optimization of the 3D matching problem involving three disjoint sets (i.e., sources, IRSs, and destinations). Thus, our MINLP problem in~\eqref{eq_opt_prob} is NP-hard since it combines optimizing over discrete variables with the challenge of dealing with nonlinear functions, and therefore, its solution usually involves an enormous search space with exponential time complexity.
\subsubsection{\textbf{Problem Reformulation}}
Every NP-complete problem can be solved by the ES method. However, when the size of the instances grows, the running time becomes forbiddingly large, even for instances of fairly small size.
Therefore, we are motivated to design an efficient solution.
Specifically, we first decompose the 3D association problem into 2D sub-problems and reformulate the association optimization problem. Let $\Xi_{\tx_k,\IRS_n , :}$ and $\Xi_{{\tx_{k^\star}},{\IRS_{n^\star}},{\rx_l}}$ represent the SINRs from $\tx_k\to \IRS_n$ and $\tx_{k^\star}\to \IRS_{n^\star}\to \ue_l$, respectively. Here, the symbol ``$:$'' implies that the third dimension of $\vec{\Omega}$ is not taken into account. Since the IRS is neither a destination nor a source, $\Xi_{\tx_k,\IRS_n, :}$ and $\Xi_{{\tx_{k^\star}},{\IRS_{n^\star}},{\rx_l}}$ are pseudo quantities estimated based on the Channel state information (CSI) of the network. The original problem can be decomposed into two sub-problems, where the obtained solutions are sub-optimal solutions to problem~\eqref{eq_opt_prob}.

\subsubsection*{S-I Matching Problem}
From the pseudo quantity, we define the following pseudo data rate of the channels 
\begin{align}\label{ratekn} 
    \Lambda_{\tx_k,\IRS_n , :}  \triangleq \log_2(1+ \Xi_{\tx_k,\IRS_n , :}).
\end{align}

Henceforth, we define $[\boldsymbol{\Lambda}]_{\tx_k, \IRS_n}$ as a pseudo quantity matrix of the $\rx$-$\IRS$ channels, in which $\Lambda_{\tx_k,\IRS_n , :}$ denotes its $(k, n)$th element.
Using \eqref{ratekn}, the first sub-problem of the decomposed problem can be expressed as 
\begin{subequations}
\begin{align}
& \vec{P1} \text{ : } \underset{[\vec{\Omega}]_{\tx_k,\IRS_n , :}}{\text{\small maximize}}
& 
& \textstyle  \sum_{k=1}^K  \Lambda_{\tx_k,\IRS_n , :}, \label{sp1}\\
&\quad\quad\text{subject to} 
&& \eqref{eq_original_prob_constraint_1}, \eqref{eq_original_prob_constraint_2}, \eqref{eq_original_prob_constraint_4}.
\end{align}
\end{subequations}

In this phase, we achieve the matching between $\s_{k^\star}$ and $\IRS_{n^\star}$, and the achieved association matrix can be denoted as $[\vec{\Omega}]_{\tx_{k^\star},\IRS_{n^\star} , :}$.
\subsubsection*{I-D Matching Problem}
After pairing  S-I, the association matrix $[\vec{\Omega}]_{\tx_{k^\star},\IRS_{n^\star} , :}$ is utilized  and considered as $\tx_{k^\star}$ is paired with $\IRS_{n^\star}$. Consequently, we formulate the pseudo data rate of the I-D as follows 
\begin{align} \label{ratenl}
  &  \Lambda_{\tx_{k^\star},\IRS_{n^\star}, \rx_l}  \triangleq \log_2(1+ \Xi_{{\tx_{k^\star}},{\IRS_n^\star},{\rx_l}}).
\end{align}

Henceforth, we define $[\boldsymbol{\Lambda}]_{\IRS_{n^\star}, \rx_l}$ as a pseudo quantity matrix of the  $\IRS$-$\rx$ channels, in which $\Lambda_{\tx_{k^\star}, \IRS_{n^\star}, \rx_l}$ denotes its $(n^\star, l)$th element. Using \eqref{ratenl}, the second sub-problem can be formulated as
\begin{subequations}
\begin{align}
& \vec{P2} \text{ : } \underset{[\vec{\Omega}]_{\tx_{k^\star},\IRS_{n^\star},\rx_l}}{\text{\small maximize}}
& 
& \textstyle  \sum_{l=1}^{L}  \Lambda_{\tx_{k^\star},\IRS_{n^\star},\rx_l}, \label{sp2}\\
&\quad\quad\text{subject to} 
&& \eqref{eq_original_prob_constraint_2}, \eqref{eq_original_prob_constraint_3}, \eqref{eq_original_prob_constraint_4}.
\end{align}
\end{subequations}
\section{Proposed Matching-based Solution} \label{proposed}
IRS-aided networks require a massive number of reflecting elements $M$, especially at higher frequencies. In this context, our model contains multiple distributed IRSs with a large number of reflecting elements. Thus, the ES method is not an efficient scheme, even for 2D association problems. Therefore, to optimize the 2D association problems in $\vec{P1}$ and $\vec{P2}$, we propose a Gale-Shapley algorithm based~\cite{gale1962college} solution, presented in the following sections.

\begin{algorithm}[t]
  \caption{Proposed Algorithm for $\vec{P1}$ and $\vec{P2}$}
  \label{algo:1}
  \DontPrintSemicolon{
  \KwIn{ $K$, $N$, $L$, responders data rate $\Lambda_{r,p}$, pseudo quantity matrix $[\boldsymbol{\Lambda}]_{p, r}$, set of unmatched proposers $\Pi$, association matrix $[\vec{\Omega}]_{p,r}$, and the set of stable matched pairs ${\cal S}$.}
 {\bf Initialize } $[\vec{\Omega}]_{p,r} = \boldsymbol{0}^{|{\cal P}| \times |{\cal R}|}$ and 
 ${\cal S} = {\tt \emptyset}$.\;
\underline{\textbf{$\vec{P1}$ : S-I Association:}}\\ \label{p1s}
$K$ sources as proposers and $N$ IRSs as responders so that $\mathcal{P} = \{ \tx_1, \tx_2, \dots, \tx_K \}$ and $\mathcal{R} = \{\IRS_1, \IRS_2, \dots, \IRS_N \}$.\;
Follow the matching algorithm in steps~\ref{ma} to~\ref{ed}. \;
{\bf Output of P1 :} S-I association matrix $[\vec{\Omega}]_{\tx_{k^\star},\IRS_{n^\star} , :}$. \label{p1e}

\underline{\textbf{$\vec{P2}$ : D-I Association:}}\\ \label{p2s}
Output of $\vec{P1}$ consider as a responders, destinations as proposers, and $K=L$ so that $\mathcal{P} = \{ \rx_1, \rx_2, \dots, \rx_L\}$ and $\mathcal{R} = \{\IRS_1, \IRS_2, \dots, \IRS_N \}$. \;
Follow the matching algorithm in steps~\ref{ma} to~\ref{ed}. \;
{\bf Output of P2 : } (S-I)-D association matrix $[\vec{\Omega}]_{\tx_k,\IRS_n,\rx_l}^\star$. \label{p2e}
\underline{\textbf{Priority Matrix Configuration:}} 

Sort each row of $[\boldsymbol{\Lambda}]_{p, r}$ and $[\boldsymbol{\Lambda}]_{r, p}^{\sf T}$ to obtain the priority matrices as\;
$[\sim, [\boldsymbol{\Upsilon}]_{p, r}] = \mathtt{sort}([\boldsymbol{\Lambda}]_{p, r}, 2, `\mathtt{descend}`)$. \;
$[\sim, [\boldsymbol{\Upsilon}]_{r, p}] = \mathtt{sort}([\boldsymbol{\Lambda}]_{r, p}^{\sf T}, 2, `\mathtt{descend}`)$.
\underline{\textbf{Matching Algorithm:}} \\ 
{\bf Initialize }$ \Pi = \mathcal{P}$ and ${\cal R}_p = \mathcal{R}$, $\forall p \in \mathcal{P}$, where $\mathcal{R}_p$ is the set of available responders of proposer $p$. \; 
\While {(either $\Pi$ $\neq \emptyset$ or proposers not rejected by all responders) \label{ma}}{
\For {$p^\star \in \Pi$}{
    Get $r^\star = \max_{r \in \mathcal{R}_{p^\star}} [\vec{\Upsilon}]_{p^\star, r}$ as the current responder with the highest priority. \;
    Update $\mathcal{R}_{p^\star} \leftarrow \mathcal{R}_{p^\star} \backslash \{ r^\star \}$. \; 
    \If{${\sum}_{p \in \mathcal{P}} [\boldsymbol{\Omega}]_{p, r^\star} = 1$, i.e., responder $r^\star$ is already paired with proposer $p'$}{ 
        Get $p' = \operatorname{arg}\max_{p\in\mathcal{P}}~[\boldsymbol{\Omega}]_{p, r^\star}$ as the paired proposer of the responder $r^\star$. \;
        \If{$[\boldsymbol{\Upsilon}]_{r^\star, p^\star} > [\boldsymbol{\Upsilon}]_{r^\star, p'}$,
        i.e, the responder $r^\star$ prefers proposer $p^\star$ than proposer $p'$}{
            $[\boldsymbol{\Omega}]_{p', r^\star} = 0$ and $[\boldsymbol{\Omega}]_{p^\star, r^\star} = 1$, \;
            ${\cal S} \leftarrow [{\cal S} \backslash \{(p', r^\star)\}] \cup \{(p^\star, r^\star)\}$, \; 
            $\Pi \leftarrow [\Pi \backslash \{ p' \}] \cup \{ p^\star \}$. \; 
        }
    }
    \Else{
        $[\boldsymbol{\Omega}]_{p^\star, r^\star} = 1$ and ${\cal S} \leftarrow {\cal S} \cup \{(p^\star, r^\star)\}$, \;
        $\Pi \leftarrow \Pi \backslash \{ p^\star \}$. \;
    }\label{ed}
} 
}
}
\end{algorithm}

\subsubsection{\textbf{$\vec{P1} \text{ : }$Proposed S-I Matching}} \label{proposal1}
The first sub-problem, i.e., $\vec{P1}$, seeks to maximize a pseudo quantity between S-I link through S-I pairing. The pseudo quantity from $\tx_k$ to $\IRS_n$ is given in~\eqref{ratekn}. Upon calculating the pseudo quantity matrix $[\boldsymbol{\Lambda}]_{\tx_k, \IRS_n}$, the source priority matrix $[\boldsymbol{\Upsilon}]_{\tx_k, \IRS_n}$ is then constructed. 
    Furthermore, the pseudo quantity matrix from the IRSs to the sources is defined as ${[\boldsymbol{\Lambda}]_{\IRS_n, \tx_k} \triangleq [{\boldsymbol\Lambda}]_{\tx_k, \IRS_n}^{\sf T}}$. 
The priority matrix at sources/IRSs is constructed with the highest pseudo quantity offered by the IRSs/sources at the top and the lowest pseudo quantity offered by the one at the bottom. We consider the sources as proposers and IRSs as responders and obtained the association matrix, as shown in steps \ref{p1s} to \ref{p1e} of Algorithm~\ref{algo:1}. Furthermore, the matching algorithm is explained in Section~\ref{pm}.

\subsubsection{\textbf{$\vec{P2} \text{ : }$Proposed (S, I)-D Matching }} \label{irsu}
The association matrix obtained in $\vec{P1}$ can be used to achieve the I-D association matrix. In this way, we achieve the S-I-D association matrix as a sub-optimal solution. Considering that ($\tx_{k^\star}, \IRS_{n^\star}$) is the pair obtained in the S-I association, the data rate offered by $\ue_l$ to the ($\tx_{k^\star}, \IRS_{n^\star}$) channel is given in~\eqref{ratenl}. 
    Combining all the data rates between IRSs and destinations yields a pseudo data rate matrix, $[\boldsymbol{\Lambda}]_{\IRS_{n^\star}, \rx_l}$, and the pseudo date rate matrix from all the destinations to all the IRSs is defined~as
    $[\boldsymbol{\Lambda}]_{\rx_l, \IRS_{n^\star}} \triangleq [\boldsymbol{\Lambda}]_{\IRS_{n^\star}, \rx_l}^{\sf T}$. 
In this context, the priority matrix is created in descending order of priorities. 
    Accordingly, the destination/IRS with the highest priority comes first for each IRS/destination in the priority matrix.  
The priority matrix for IRSs and destinations are denoted as $[\vec{\Upsilon}]_{\IRS_{n^\star}, \rx_l}$ and $[\vec{\Upsilon}]_{\rx_l, \IRS_{n^\star}}$, respectively. It is noted that $[\vec{\Upsilon}]_{\IRS_{n^\star}, \rx_l} \ne [\vec{\Upsilon}]_{\rx_l, \IRS_{n^\star}}^{\sf T}$.
    We consider the destinations as proposers and IRSs (i.e., the IRSs allocated in $\vec{P1}$) as responders and obtained the association matrix, as shown in steps \ref{p2s} to \ref{p2e} of Algorithm~\ref{algo:1}, and the matching algorithm is explained in Section~\ref{pm}. In this way, we can achieve the near-optimal solution of $\vec{P}$.
\subsubsection{\textbf{Proposed Matching Algorithm}} \label{pm}
We explore proposer-responder ($p$-$r$) association algorithms as one-to-one matching. Each proposer can be allocated to at most one responder in one-to-one matching. The remaining responders, which are not allocated to any proposer, are considered inactive for this round. The algorithm, shown in steps~\ref{ma} to~\ref{ed} of Algorithm~\ref{algo:1}, is illustrated as a series of attempts from proposers to responders. The responder can either be associated with the proposer or remain vacant during this process. Each proposer proposes pairing with the highest priority responder in their priority list until the proposer has either been paired or rejected by all responders. Proposers rejected by the responder are not allowed to try again for the same responder. The responder is immediately engaged if the proposer proposes a free responder; however, if the proposer proposes the responder that is already engaged, the responder compares the new and current proposer and gets engaged with the best one, i.e., the one with the higher pseudo quantity link. The process is repeated until all proposers are engaged or all options are considered.
\begin{Lemma} \label{theorem1a}
The proposed \textit{\textbf{proposer}}-\textit{\textbf{responder}} association algorithm terminates in polynomial time, i.e., if there are $K$ \textit{\textbf{proposers}} and $N$ \textit{\textbf{responders}}, then the algorithm is terminated after at most $K \times N$ iterations.
\end{Lemma}
\begin{IEEEproof}
In each iteration, each unmatched \textit{\textbf{proposer}} proposes to a \textit{\textbf{responder}} that it has never explored before. Moreover, the \textit{\textbf{responder}} accepts or rejects the new proposal according to the status and preference of the \textit{\textbf{responder}}. As a result, for $N$ \textit{\textbf{responders}} and $K$ \textit{\textbf{proposers}}, we have $K \times N$ possible proposals occurring in the proposed algorithm. This completes the proof of Lemma \ref{theorem1a}.
\end{IEEEproof}
\begin{Theorem} \label{theorem2}
The \textit{\textbf{proposer}}-\textit{\textbf{responder}}  matching achieved for each phase of the proposed algorithm returns a stable matching. Furthermore, the stability is not affected by inverting the proposing order.
\end{Theorem}
\begin{IEEEproof}
We prove that there is a stable matching obtained by the proposed algorithm using the contradiction as follows. 
Meanwhile, stability means that each proposer making a proposal is matched with the responder, which is most preferred by the proposer among all available options.
    Suppose that some \textit{\textbf{proposer}} $p \in \mathcal{P}$ is rejected by the best valid \textit{\textbf{responder}} $r=best(p)$ ($r \in \mathcal{R}$) in the proposed algorithm. 
Furthermore, $r$ rejects $p$ in favor of $p^\star \in \mathcal{P}$, which $r$ prefers $p^\star$ than $p$. Thus, in $r's$ priority, we have $p^\star \succ^{{r}} p$. 
    Let us indicate this point as $\textit{\textbf{R}}$ (rejection point), and return to it later to drive a contradiction. According to the hypothesis that $r=best(p)$, there exists an edge between vertex $p$ and $r$ in $G$, which makes a stable matching $\Omega'$. 
Moreover, in $\Omega'$, there is an edge between $p^\star$ and $r^\star \in R$ in a bipartite graph $G$. 
Thus, $\Omega'$ contains stable matches ($p, r$) and ($p^{\star}, r^\star$). 
    We now consider what the implementation of the proposed algorithms decides about the priorities of $p^\star$ between $r$ and $r^\star$. Since $\textit{\textbf{R}}$ was the first event in the algorithm where any \textit{\textbf{proposer}} was rejected by its best valid partner, at this instant, the following must be true: $(i)$ $p^\star$ has not been rejected by its $best(p^\star)$, and  $(ii)$ $p^\star$ has been rejected by every \textit{\textbf{responder}} in the list that comes before $r$, since $p^\star$ is paired with $r$ at instant $\textit{\textbf{R}}$.
The above two facts jointly imply that $r \succ^{{p^\star}} r^\star$ in the priority list of $p^\star$. First, $best(p^\star)$ must come after $r$ because prior hasn't rejected $p^\star$ yet, so $r > best(p^\star)$. Secondly, $r^\star$ is always a valid match with $p^\star$ because $\Omega'$ is a stable matching. This means ($p^\star, r$) is an unstable match for $\Omega'$; however, in $\Omega'$, both $p^\star$ and $r$ prefer each other to their assigned pairs. Thus, our initial assumption that $p$ was rejected by $best(p)$ is false, which proves that the algorithm returns the stable matching.

Now, we prove that the stability does not change when inverting the proposing order (i.e., previous \textit{\textbf{proposer}} as \textit{\textbf{responder}} and \textit{\textbf{responder}} as \textit{\textbf{proposer}}). In our proposed work, we consider the channel between the source and the IRS in the first phase, while in the second phase, the channel between the IRS and the destination is considered. Moreover, channel reciprocity holds for the S-I and I-D channels. Thus, inverting the proposing side doesn't affect the stability of the proposed algorithms. This completes the proof of Theorem 1.
\end{IEEEproof}
\section{Results and Discussion} \label{pref} 
In the simulated scenario, explained in Section~\ref{model}, the sources with transmit power of 25 dBm, destinations, and IRSs are deployed within the network area. The length and width of each IRS element is $0.4 \lambda$~\cite{najafi2020physics}, where $\lambda$ is the wavelength. Additionally, the transmission carrier frequency of 300 GHz, the bandwidth of 10 GHz, absorption coefficient 0.0033 $m^{-1}$, the noise power density of -174 dBm/Hz, and noise figure 10 dB are considered in the simulation based on existing works. Furthermore, the proposed scheme is compared to the following schemes: 1) In the ES method, the S-I-D sum rate is utilized to find the optimal pairing by checking all possible combinations and selecting the optimal. 2) In the partial ES (PES) scheme, all the possible S-I pairing combinations are explored. Following this association, PES explores the optimal I-D matching, which leads to sub-optimal S-I-D matching. 3) In the GS method, each source/destination selects/proposes the IRS with a higher data rate. IRSs receiving multiple proposals are randomly allocated to one of the sources/destinations. 3) The RA scheme randomly selects the S-I-D association matrix. 4) The partial RA (PRA) scheme randomly selects the S-I association matrix, and then also randomly selects the I-D association matrix. 
 \begin{figure}[t]
 \centering
 \includegraphics[width=\linewidth]{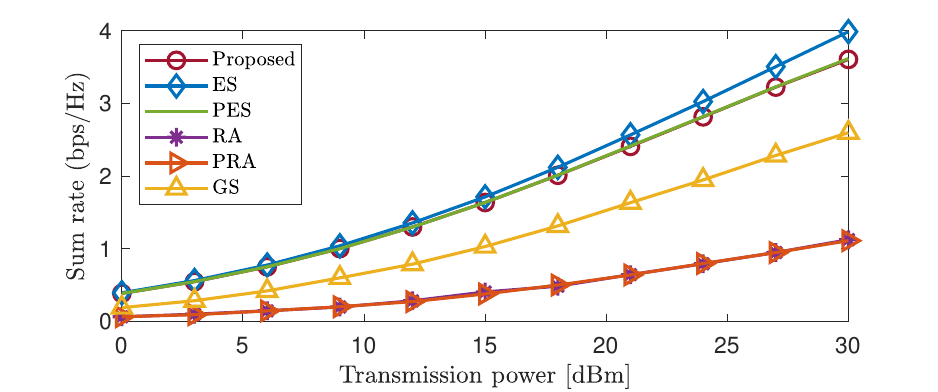}
 \caption{Sum rate versus transmit power for a network area of $20\times20$ $\mathrm{m}^2$ and $M=100\times100$.}
 \label{sr_p}
 \end{figure}
For all cases, the sum rate increases with the transmission power, as shown in Fig.~\ref{sr_p}. Additionally, the sum rate of the proposed scheme matches that of the PES scheme and is comparable to that of the ES scheme. Furthermore, the proposed algorithm outperforms the GS, PRA, and RA schemes.

\begin{figure}[t]
 \centering
 \includegraphics[width=\linewidth]{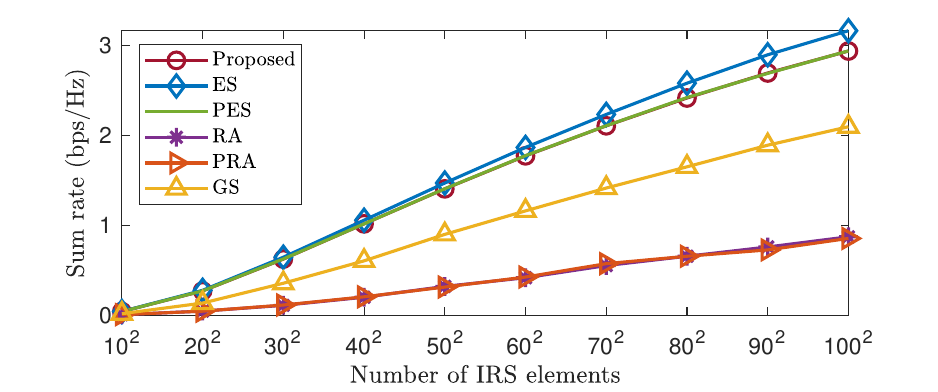}
 \caption{Sum rate versus number of IRS elements for a network area of $20\times20$ $\mathrm{m}^2$ and $P_{k} =25$ dBm.}
 \label{sr_m}
 \end{figure}

Fig.~\ref{sr_m} demonstrates that the sum rate increases with $M$. This is because the received signal strength increases with $M$, which ultimately improves the sum rate. Moreover, the sum rate of the proposed algorithm is similar to that of the PES scheme and comparable to the ES method. Furthermore, the proposed scheme achieves a higher sum rate than the GS, RA, and PRA algorithms, which is due to proper association.
\section{Conclusions}\label{conc}
In this paper, we considered IRS-aided green wireless communications for THz networks where several IRSs are deployed to assist the communication. First, we established the PL model for the S-I-D cascaded channel. Then, we formulated the S-I-D matching problem with the aim of maximizing the sum rate, which is an NP-hard problem. Thus, to address the NP-hard problem, we divided the problem into two sub-problems and employed a matching-based algorithm to solve them. Simulation results demonstrated the effectiveness of our proposed scheme for IRS-aided THz communications, outperforming existing GS, PRA, and RA schemes. 

\bibliographystyle{IEEEtran}
\bibliography{main}
\end{document}